**Corresponding Author:**
Richard Ashworth, Airbus Group Innovations, Building 20A1, Bristol, BS34 7QW, United Kingdom
Email: richard.ashworth@airbus.com

# Stabilization of Crossflow Instability with Plasma Actuators: Linearized Navier Stokes Simulations

Kean Lee Kang[1]. Richard Ashworth[1]. Shahid Mughal[2]

[1]Airbus Group Innovations, Building 20A1, Bristol, BS34 7QW, United Kingdom

[2]Department of Mathematics, Imperial College, London, SW7 2AZ, United Kingdom

## Abstract

This paper describes work carried out within the European Union (EU)-Russia Buterfli project to look at the control of transition-causing "target" stationary cross flow vortices, by the use of distributed plasma actuation to generate sub-dominant "killer" modes. The objective is to use the "killer" modes to control the "target" modes through a non-linear stabilizing mechanism. The numerical modelling and results are compared to experimental studies performed at the TsAGI T124 tunnel for a swept plate subject to a favorable pressure gradient flow. A mathematical model for the actuator developed at TsAGI was implemented in a linearized Navier Stokes (LNS) solver and used to model and hence predict "killer" mode amplitudes at a measurement plane in the experiment. The LNS analysis shows good agreement with experiment, and the results are used as input for non-linear PSE analysis to predict the effect of these modes on crossflow transition. Whilst the numerical model indicates a delay in transition, experimental results indicated an advance in transition rather than delay. This was determined to be due to actuator induced unsteadiness arising in the experiment, resulting in the generation of travelling crossflow disturbances which tended to obscure and thus dominate the plasma stabilized stationary disturbances.

## Keywords

Flow control, plasma actuator, laminar turbulent transition, crossflow instability, linearized Navier Stokes, parabolized stability equations, PSE

## Introduction

The laminar-turbulent transition process over a swept transport wing under cruise conditions is dominated near the leading edge by crossflow instability. A common approach to controlling such disturbances in laminar wing design involves modifying the base flow of the boundary layer through suction to make it less susceptible to the growth of these instabilities. The system to achieve this incurs a drag penalty due to weight and actuation energy and raises reliability issues due to the complex network of pipes and pumps required. An alternative approach for controlling laminar-turbulent transition at the leading edge is through the promotion of "killer" crossflow modes that retard the growth of the most unstable "target" transition-causing modes via nonlinear interaction. This can be realized through excitation of the "killer" modes with distributed roughness elements or through an active approach involving spanwise periodic forcing, for example with plasmas, heat spots or blowing. Such active approaches for introducing disturbances require less energy compared to approaches involving base flow modification because they involve only small perturbations to the flow over a much more limited chord-wise extent. Furthermore, such a dynamic method offers the possibility of varying the wavelength of the control perturbations to match the changing stability properties of the boundary layer under variations in cruise conditions that occur during a typical aircraft flight. Plasma actuators appear to be well suited to this role, offering the benefit of rapid response time and flexibility and are capable of producing the small velocity amplitudes that are required.

The viability of this approach has been explored in the European Union (EU)-Russia "Buterfli" project. In this project a distributed dielectric barrier discharge (DBD) plasma actuator was designed and built by the Joint Institute for High Temperatures (JIHT) in Russia and incorporated into a swept plate model at TsAGI (Ustinov et al.[1]) for testing in the T124 low turbulence tunnel under conditions where transition occurs due to crossflow instability. The effectiveness of the control through plasma actuation has been explored through the tunnel test and also through numerical simulations, the results of which are reported in this paper. This involves implementation of an actuator body force model developed by TsAGI and JIHT, as a field source term in a Linearized Navier Stokes (LNS) code (Mughal[2]). Initial "killer" mode amplitudes determined with the LNS solver are used to set upstream "linear regime" conditions for the nonlinear parabolized stability equations (PSE, see Herbert[3]). Through solution of the PSE, the growth of the primary "killer" crossflow disturbance growth along with the transition-causing mode and their nonlinear interactions are studied to determine the possible transition delay achievable downstream. Although the predicted "killer" modes' evolution match experimental downstream plane measurements, the predicted delay in transition from the numerical study does not correspond with experiment due to additional unsteady effects arising from the plasma actuation device.

The DBD actuator creates a wall jet flow effect, which with careful tuning of the plasma force distribution, can be confined to cause maximum effect within the viscous boundary layer. The wall jet arises with the introduction of a body force in the fluid whereby the plasma actuator enhances the local fluid momentum (Riherd & Roy[4]). Plasma actuation produces a very concentrated and thus localized flow modification and beyond the DBD placement region, the plasma induced force reduces in magnitude fairly rapidly. Such extreme modifications to the basic field have been modelled until now by solutions to non-linear boundary layer, Direct Navier Stokes (DNS) or Reynolds Averaged Navier-Stokes (RANS) equations.

The physics of plasma actuation is a very complex problem, due to the multi-time and multi-spatial scales arising and in general requires precise numerical modelling of the convection-diffusion dynamics with the added complication of stiffness in the plasma reaction/ionization equations. This multi spatial and time modelling problem involves many stages (note Table 1, taken from Boeuf et al.,[5] see also Regis[6]). The detailed analysis of Vidmar & Stalder[7] indicates electron temperatures ranging between 1000K to 10,000$K$, with electron velocities of the order of $10^5$ to $10^6$$m/s$ (Mertz & Corke[8]), while fluid velocity is typically of order of 100$m/s$. Complete numerical solutions of the fully coupled plasma-fluid dynamics in a consistent manner, in addition to Vidmar & Stalder,[7] are the works of Unfer & Boeuf,[9] Singh & Roy[10] and Orlov[11] among others.

Table 1: Scales involved in DBD actuator modelling (Boeuf et al.[5], more detailed analysis of the disparate time and spatial scales is in Unfer & Boeuf [9])

| Temporal scales | seconds |
|---|---|
| Maxwell relaxation time | $10^{-12}$ |
| CFL time - electrons | $10^{-12}$ |
| CFL time - ions | $10^{-10}$ |
| Plasma formation | $10^{-9}$ |
| Voltage generator | $10^{-4}$ |
| Ambipolar diffusion | $10^{-4}$ |
| Generated flow | $10^{-2}$ |
| Spatial scales | metres |
| Sheath | $10^{-6}$ |
| Plasma dimension | $10^{-3}$ |
| CFL | $10^{-1}$ |
| Generated flow | $10^{-1}$ |

An alternative route has been that based on the electrostatic body force approximation (proposed by Shyy, Jayaraman & Andersson[12]) based on the key assumption, as alluded to above, that the many processes involved in the physics of the actuator operation (such as charge re-distribution) occur across widely separated time and spatial scales, so that the plasma generation may be treated in a quasi-steady manner. If one could directly implement an adequate approximation of the electrodynamic force in the Navier-Stokes equations, significant computational cost savings could be obtained for solutions of the complete plasma-fluid coupled equations.

The modelling challenge is thus that of developing a body-force model that captures the true induced force produced by the plasma hardware under flight or experimental operating conditions. A common practice is that of making measurements of the plasma actuator force vector under quiescent conditions and then applying the experimentally captured data in non-quiescent test conditions (see Serpieri et al.[13]).

The plasma actuator body force $f$, based on experiments and comparison with full and detailed numerical simulations of the plasma generation process, was found to be well approximated by Maxwell's electrostatic force equations. Mertz & Corke,[8] Orlov[11] and others solve the Poisson equation numerically to compute the electric potential distribution and thus the body force components. Shyy, Jayaraman & Andersson[12] proposed a simple empirical model to characterise the DBD actuator, with the assumption that the plasma only arises above the actuator, in a triangular shaped region with a linear decrease of the field strength from the maximum point; the field strength being maximal at the minimum separation between electrodes. The effect of increasing voltage was then modelled by a re-scaling of the triangular domain with voltage increments. This simple model has been used in a number of Euler, DNS

and RANS solvers (Jayaraman & Shyy[14]) to corroborate the flow control effects arising from the plasma actuator and with experimental verification. This simplistic expression of course, requires very fine tuning and fitting to experiment, otherwise the total DBD force acting on the flow is erroneous at higher, and *off design* voltages; giving rise to negative and unphysical force components. Singh & Roy[10] and Grundmann, Klumpp & Tropea[15] build upon this approach, by proposing a considerably more sophisticated and highly tuneable induced body force vector. The basis of their model is detailed multiscale modelling of a two-dimensional DBD with a finite element time adaptive method. This detailed modelling confirms that the maximum plasma force arises in the overlap between the exposed and grounded electrodes. Assuming that an essentially time averaged body force arises, and coupled to choosing initial conditions matched with realistic experimental data they proposed the electrodynamic force could be well approximated by exponential forms. Generally they find that for fixed frequency, the body force induced jet velocity increases with the applied voltage.

Their body force expression is of course strictly two-dimensional, an accurate description of the electric field that is strongly tied to the electrode 3D-shape is a much more difficult problem to model and parameterise, which will be case and shape dependent. Disotell[16] considers the implication of having a periodic array of shaped plasma actuators and studies them in a semi-empirical manner. More recent works concerned with controlling flow instabilities as opposed to flow separation are those of Wang, Wang & Fu[17] (with a nonlinear PSE plasma model) and Dorr & Kloker[18,19,20] (compute and resource intensive DNS modelling) . In addition to Ustinov et al.,[1] whose experiment we replicate in this paper numerically, experimental plasma control of crossflow instability using the killer mode concept has also been reported by Serpieri, Venkata & Kotsonis.[13]

In the PSE model of Wang, Wang & Fu[17] a sensitivity analysis is undertaken of the effect of plasma on the nonlinear crossflow arising in the swept Hiemenz base flow. Due to the inability of the PSE to model the disturbance generation process arising from the plasma actuation itself, they model how a pre-existing crossflow disturbance state (generated by some other means) is modified as it convects through the plasma (*i.e.* a sensitivity analysis) region. In addition to this simplification there is also the assumption of a very weak plasma momentum force, to enable the nonlinear PSE (NPSE) study to be feasible.

We investigate the feasibility of modelling plasma flow control through the linearized Navier Stokes model. The plasma actuation within the LNS scheme of modelling the periodic spanwise actuation of stationary crossflow vortices is of an exploratory but novel nature. We add in the three-dimensional actuation capability by a smoothly varying step function (in the *y*-spanwise direction) to give periodic variations in span for the body-force source terms. Hence, the spanwise variation is simply a smoothed step up/down function (ranging between 0 and 1) which allows the plasma to be active in a specified fraction of the periodic box under investigation along similar lines as that shown in Figure 1. As remarked above most work in plasma actuation has been based on making significant alterations to the flow, for example to control flow separation. The linearised Navier Stokes model herein will be a valid model for magnitudes of the body-force source terms which lead to weak changes to the leading order basic field. Unlike using the full *brute-force* approach of Dorr & Kloker,[18] the harmonic LNS route affords an efficient means of investigation of this relatively weak phenomenon. Not only does the LNS method unlike the PSE model, capture the birth process of crossflow disturbance due to the plasma, non-parallel and any detailed short-scale physical processes arising in the disturbance evolution are also captured. PSE models cannot model processes below a certain spatial resolution due to an inherent ill-posedness in the PSE equations and hence step-size

restriction. This has been demonstrated in the context of surface roughness induced stationary crossflow generation by Mughal & Ashworth.[21] All of the LNS results reported in this paper were undertaken on a stand-alone workstation, with each LNS simulation taking less than 5 minutes for a highly resolved grid-independent computation.

The details of our numerical approach may be found in the paper of Thomas, Mughal & Ashworth,[22] below we focus on aspects of the plasma model and demonstrate the efficacy of our LNS model with comparisons with the experiment. The basis of our methodology is as follows: we utilise the LNS model for capturing the plasma induced crossflow generation of the killer mode; and the amplitude of the killer mode predicted by the LNS model is then used to force the significantly more efficient nonlinear PSE solver to investigate the control aspects of the study.

# Experiment

The experiment in the TsAGI T124 tunnel has been described in detail in Ustinov et al.[1,23] In summary a flat plate with 35 degree sweep was subjected to a favorable pressure gradient flow on one surface through tunnel wall profiling. This was designed to promote the development of cross flow instabilities. Hot wire velocity profiles measurements were made across three spanwise planes at distances of 250 *mm*, 350 *mm* and 600 *mm* from the leading edge. The last of these was expected to be close to the onset of transition. Extensive surface pressure data was also collected. Preliminary numerical studies[23,24] had shown that the most amplified naturally occurring stationary cross flow mode was close to 7.5 *mm* in spanwise wavelength and this could most effectively be controlled through introduction of a "killer" mode having 2/3 of the spanwise wavelength *i.e.* 5 *mm*. Thus a DBD actuator was developed by JIHT to promote this particular mode through a periodic arrangement of electrodes (Moralev et al.[25]). This consisted of two embedded electrodes as shown in Figure 1. The upper embedded electrode included equidistantly spaced electrodes at the actuation wavelength ($\lambda$=5 *mm*) and an electrode width (*w*=2 *mm*). The DBD actuator was embedded in the plate with the downstream edge of the exposed electrode at a distance of 120 *mm* from the leading edge of the plate.

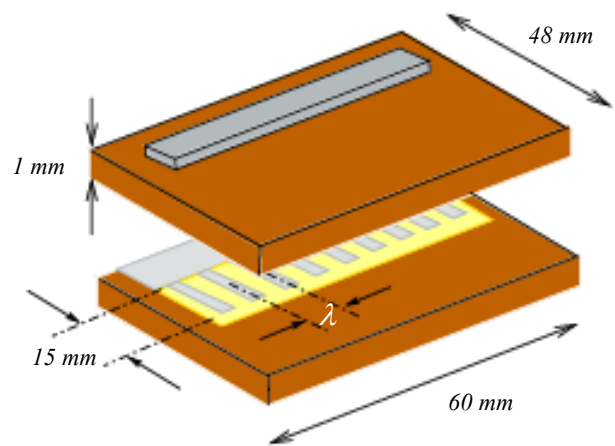

Figure 1: Sandwich actuator with control electrodes (JIHT report by Moralev et al.[25])

Measurement data was collected with and without actuation at a tunnel speed of 25 *m/s* which was equivalent to a virtual freestream of 31.9 *m/s*. The plasma actuation in the experiment, unlike the simulations reported in this paper, was found to advance rather than delay transition in the experiment. From the spectral velocity data this was thought to be due to unsteady forcing

by the plasma actuator which generated travelling crossflow modes alongside the stationary modes. The travelling modes were sufficiently unstable to trigger the early transition. Finding a way of eliminating this unsteadiness is one of the major technical challenges to be overcome if plasma actuators are to be used in this way. This will be the subject of future work. This paper describes numerical modelling of the actuators neglecting the unsteady effects and certainly the simulations indicate that the actuators should be able to produce a delay in transition, should it prove possible to eliminate and manufacture plasma actuators exhibiting reduced levels of unsteadiness during operation. Nevertheless, the numerical model of the actuator arising from the Buterfli project and our simulations have been successful in validating some aspects of the experiment, and indeed we find that plasma actuators can be used to control stationary crossflow vortices.

# Numerical Model

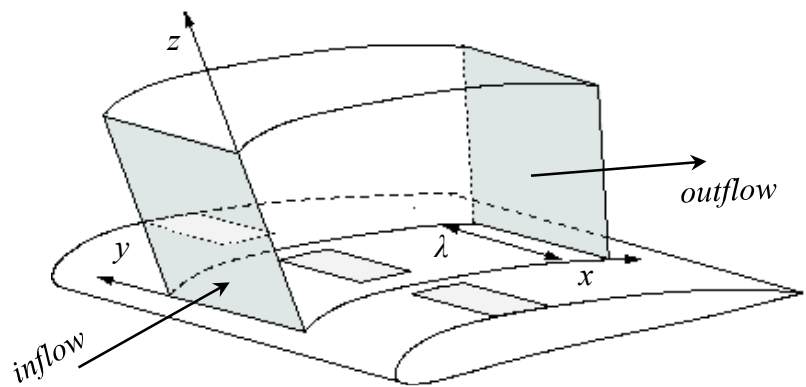

Figure 2: Computational domain of linearized Navier Stokes solver for spanwise periodic roughness or body forcing in spanwise homogeneous flow.

## Equations

The generation of the killer cross flow mode has been modelled through incorporation of a modified form of the TsAGI body force model[26] into a linearized Navier Stokes (LNS) solver developed at Imperial College with support from Airbus Group Innovations.[2] As described in some detail in Mughal & Ashworth[21] and Thomas, Mughal & Ashworth.[22] the LNS equations are solved for spanwise periodic disturbances in a spanwise homogeneous base flow with a computational domain that encloses one period of the disturbance inducing roughness or plasma forcing as indicated in Figure 2.

The derivation of the LNS equations follows the usual route of decomposing the total unsteady flow $\hat{q}$, along the following ansatz:

$$\hat{q}(x,y,z,t) = \bar{Q}(x,z) + \tilde{q}(x,y,z,t), \tag{1}$$

where $t$ is a time variable, $(x,y,z)$ spatial variables, $\bar{Q}$ represents the steady spanwise invariant in $y$ base flow velocity field, and $\tilde{q}$ the disturbance state involving the velocity components $(\tilde{u}, \tilde{v}, \tilde{w})$ and pressure $\tilde{p}$. The steady pressure field is assumed to be invariant with $(y,z)$ and thus $\bar{P} = \bar{P}(x)$ only. The disturbance state is then decomposed as follows

$$\tilde{q}(x,y,z,t) = \sum_{m=-N}^{m=N} q_{(m)}(x,z) e^{i(m\beta y - \omega t)}. \tag{2}$$

In the above, a time harmonic and span periodic ansatz is assumed for the disturbance $\tilde{q}$, i.e.

$$\frac{\partial \tilde{q}}{\partial t} = -i\omega q_{(m)};$$
$$\frac{\partial \tilde{q}}{\partial y} = -im\beta q_{(m)},$$
(3)

for a specified angular frequency and wavenumber pair $(\omega, m\beta)$, with $\beta = 2\pi/\lambda$. Substitution of the above into the unsteady body-fitted Navier-Stokes equations, followed by neglect of nonlinear terms leads to the LNS equations (see Appendix). As our focus is on control of steady crossflow modes, from here on we set $\omega = 0$, while the integer $m$ is used to model higher harmonics of the stationary crossflow which arise in the near field of the plasma actuation site, as well as to denote the same modes developing in our nonlinear PSE simulations (note Equation (1). For clarity from hereon we drop the suffix $m$-notation on the $q_{(m)}$ variable.

The plasma is represented by $\hat{f}_k$ field source terms in the forced LNS momentum equations, while roughness is represented through a "wall" boundary condition obtained from a Taylor expansion in the total flow (base plus disturbance) in conjunction with the requirement that no slip is satisfied where the surface of the roughness would be (*i.e.* in the interior of the flow rather than the "wall"). Thus,

$$q(x, y, 0) = -H(x, y)\frac{\partial \overline{Q}}{\partial z},$$
(4)

for a given height function $H$ and base flow velocity $\overline{Q} = (\overline{U}, \overline{V}, \overline{W})$. The discretised LNS equations ultimately requires the solution to a linear system of equations of the form

$$L\vec{q} = \vec{r},$$
(5)

where $\vec{r}$ is constructed from either the field source terms involving $\hat{f}_k$ for the plasma model developed by TsAGI, or and in combination with the roughness function $H$. The LNS operator $L$ is derived through a fourth-order accurate finite difference discretization in $x$ (the streamwise direction), a pseudo spectral discretisation in $z$ and a Fourier decomposition in $y$ the spanwise domain.

### TsAGI-JIHT body force model for distributed actuation

Within the Buterfli project a phenomenological body force model was formulated for the DBD actuator developed at JIHT. PIV measurements of the flow generated by the device were made by JIHT in quiescent conditions and the thrust generated by the device over one span period was determined through integration of the measured momentum flux on a control volume enclosing one spanwise period of the device. A value of $F_{x\Sigma} = 5.0 \times 10^{-6}$ N per period was arrived at for an actuation voltage of 3.2 *kV*, frequency of 190 *kHz* and electrode width of 2 *mm*. The forcing was observed from PIV to be restricted to a cuboidal domain immediately above the electrode of size 1.5 *mm*, 2 *mm* and 0.2 *mm* in the streamwise, spanwise and wall normal directions respectively. Determination of the spanwise component of the forcing was more problematic as the symmetry properties excluded the approach used for the streamwise component. However an estimate was made that the spanwise component was $F_{y\Sigma} = 1.2 \times 10^{-6}$ N in each half of the cuboidal domain with the force having opposite directions in each half of

the domain. The wall normal component of the force arising from the actuation was observed to be negligible compared to the wall tangential components and was therefore taken to be zero. TsAGI took this information from JIHT and combining it with insights gained from their parametric studies for canonical 2d (spanwise invariant) DBD actuators including in particular the location of force maxima, arrived at a more detailed model of the force distribution as a function of voltage. This is described in some detail in Ustinov et al.[26]

The field source terms $\hat{f}_k$ in the forced LNS (see Appendix) arise on performing a fast Fourier transform (FFT) in $y$ of expressions[26] for the longitudinal and spanwise components of the force density ($N/m^3$),

$$F_x = \frac{1}{x_0 y_0 z_0} F_{x\Sigma} \theta(\hat{x}) \frac{\hat{x}\hat{z}}{\pi^{1/2}} e^{-(\hat{x}^2 + \hat{y}^2/4 + \hat{z})};$$

$$F_y = \frac{1}{x_0 y_0 z_0} F_{y\Sigma} \theta(\hat{x}) \hat{x}\hat{y}\hat{z} e^{-(\hat{x}+\hat{y}+\hat{z})},$$

(6)

in which the non-dimensional coordinates are with respect to the location of the force maximum i.e. $\hat{x} = x/x_o$, $\hat{y} = y/y_o$, $\hat{z} = z/z_o$. The coordinates of the force maximum location depending linearly on the ratio of the applied voltage to the discharge voltage as follows:

$$x_0 = F_{x1}\left(\frac{V}{V_0} - 1\right);$$

$$y_0 = F_{y1}\left(\frac{V}{V_0} - 1\right);$$

$$z_0 = H_1 V.$$

(7)

The integrated values (for one electrode intersection) of the force density components depend quadratically on the voltage ratio:

$$F_{x\Sigma} = A_x \left(\frac{V}{V_0} - 1\right)^2;$$

$$F_{y\Sigma} = A_y \left(\frac{V}{V_0} - 1\right)^2.$$

(8)

The integration is over one side (left or right) for the $y$ component; $\theta$ is a Heaviside function to ensure forces are zero for negative $x$. The various coefficients have been empirically determined to have the following values: $A_x = 1.82 \times 10^{-5} N$, $A_y = 4.37 \times 10^{-6} N$, $F_{x1} = 1.43 mm, F_{y1} = 0.65 mm$ (changed from $F_{y1} = 0.95 mm$ in original model[26]) and $H_1 = 0.0312 mm/kV$. These parameters stretch the plasma field of influence in the streamwise and wall-normal directions, and along with the use of the Heaviside function $\theta$ affords maximum flexibility to fine-tune the plasma model with data from experiment.

The LNS simulations with the implemented model have been applied to a boundary layer corresponding to the experimental conditions for a tunnel reference velocity of $25 m/s$ as described in Ustinov et al.[1] The boundary layer or the steady base flow represented by $\bar{Q}(x,z)$ in Equation (1) was computed with the non-similar boundary layer solver CoBL (Thomas et

al.[27]), using an equivalent virtual freestream velocity of $31.9 m/s$ together with the corresponding pressure coefficient interpolated from experimental pressure measurements, shown in Figure 3.

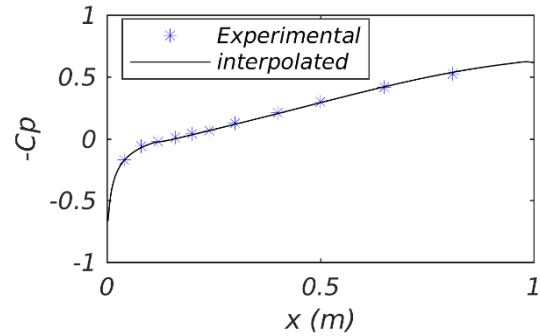

Figure 3: Pressure coefficient Cp based on virtual free stream conditions and measured data.

The interpolated pressure data was then used to compute three-dimensional boundary-layer profiles for the subsequent disturbance control simulations involving either LNS or nonlinear PSE. As a check on the correct implementation of the surface pressure data, the CoBL recomputed boundary-layer edge velocity distribution was found to compare very well with the experimentally measured edge velocities, as may be seen in Figure 4. This is a crucial check since instability analysis is known to be very sensitive to any inaccuracies or inconsistencies in the boundary-layer computations. The resulting streamwise and spanwise velocity profiles computed by our boundary-layer solver CoBL are displayed in non-dimensional form in Figure 5 at a number of streamwise positions. Validations and adequacy of the CoBL laminar boundary-layer computed solutions and comparison with RANS derived boundary layer profiles were previously reported in Thomas et al.[27]

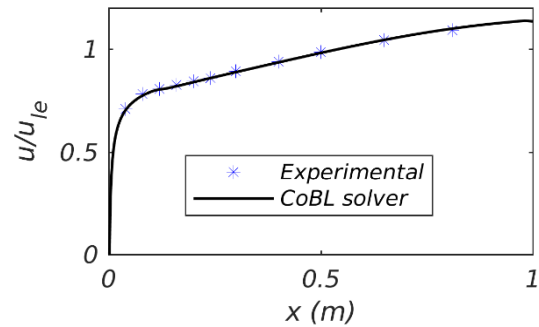

Figure 4: Comparison of computed and measured edge velocity.

## Results and Discussion

Stability analysis with a linear PSE solver for a range of stationary cross-flow modes revealed a most amplified mode of around 8 *mm*, as shown in Figure 6(a); which is in line with previous analysis of Ustinov et al.[23] and Hein.[24]

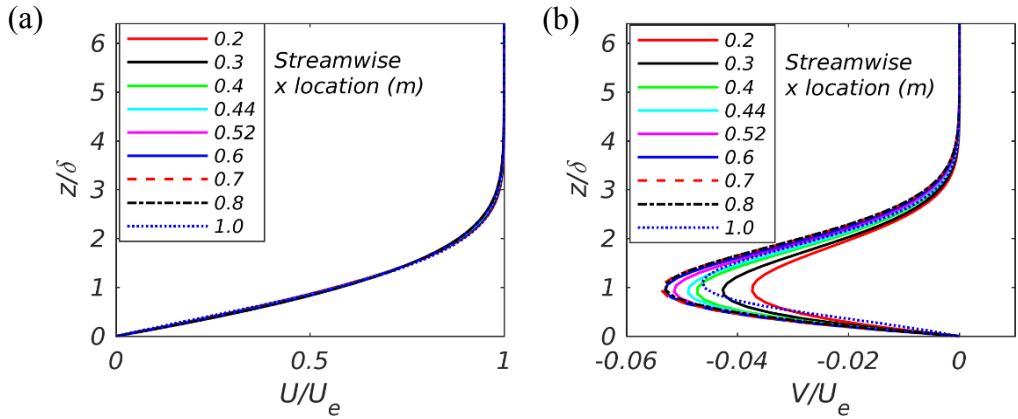

Figure 5: Non-dimensional boundary layer profiles: component in directions of inviscid stream (a) and cross-flow component (b). Velocity components and wall normal distance are normalized by local boundary layer edge velocity displacement thickness respectively.

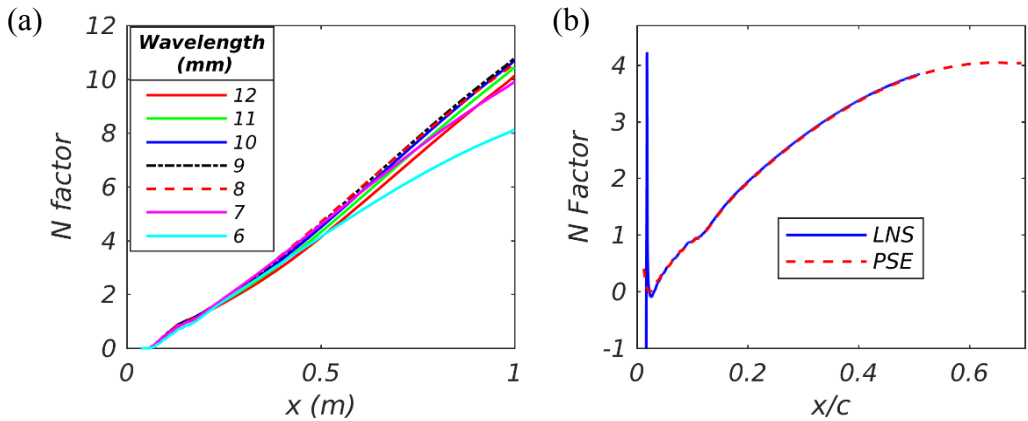

Figure 6: Linear PSE for a range of span wavelengths (a) and comparison between linear PSE and LNS simulation for the 5*mm* mode (b).

A comparison between linear PSE for the 5*mm* "killer" mode generated with our LNS simulation of a cylindrical roughness element placed within a periodic box of $\lambda = 5.0\,mm$ span showed good agreement with the amplification based N-factor, as shown in Figure 6(b). The precise details of the roughness element are unimportant, since we use it simply as a means to generate the linear stationary crossflow disturbance. The LNS solver was then applied with the plasma actuator for a range of applied voltages from $3.0\,kV$ to $4.0\,kV$ for a 5 *mm* domain spanwise width. Stationary crossflow 'killer' modes of varying magnitude are thus directly generated by the imposed varying plasma forcing. A primary concern before analysis can be conducted is to confirm the implicit assumption that the plasma body force induces only weak variations to the base flow. This is confirmed by examination of the induced steady base flow correction, the so called (0,0) mean flow distortion disturbance, computed by the LNS on setting $m = 0$ in Equation (2). Figure 7 and Figure 8(a) show the $w, v$ and $u$ components of the (0,0) mode disturbance near the plasma activation site ($x=0.11\,m$). The $u$ – component of the (0,0) mode is the largest in magnitude and indicates fluid to be accelerated in the normal to leading edge streamwise and spanwise directions whilst also being drawn in towards the surface at the commencement of the actuation (*i.e.* the relatively larger negative valued $w$ –

field in Figure 7 (a)). Observe that the plasma effect is quite localized (concentrated at about $x=0.12\ m$) for the $(w,v)$–fields, with $w$ the weakest field followed by the spanwise $v$–component a factor of ten larger, but by about $x=0.14$ all three velocity components have

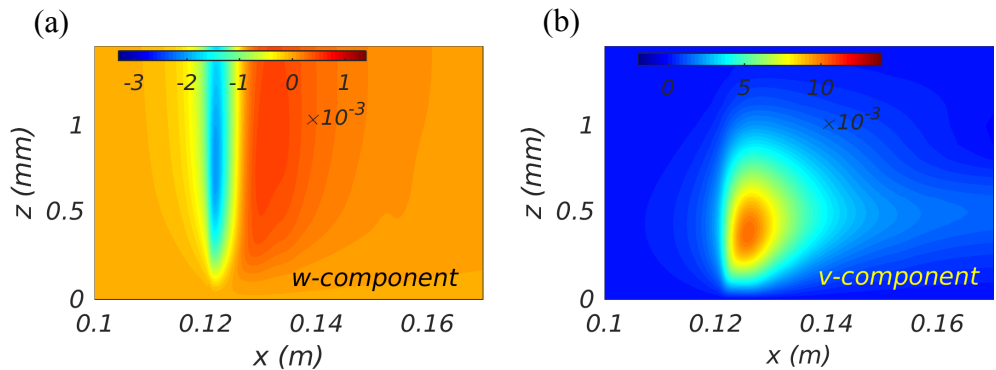

diminished in magnitude significantly.

Figure 7: Base flow modification, mean flow distortion (0, 0) mode due to plasma actuation (applied voltage 4.0 $kV$). Normal to leading edge wall normal $w$ and spanwise $v$-disturbance fields.

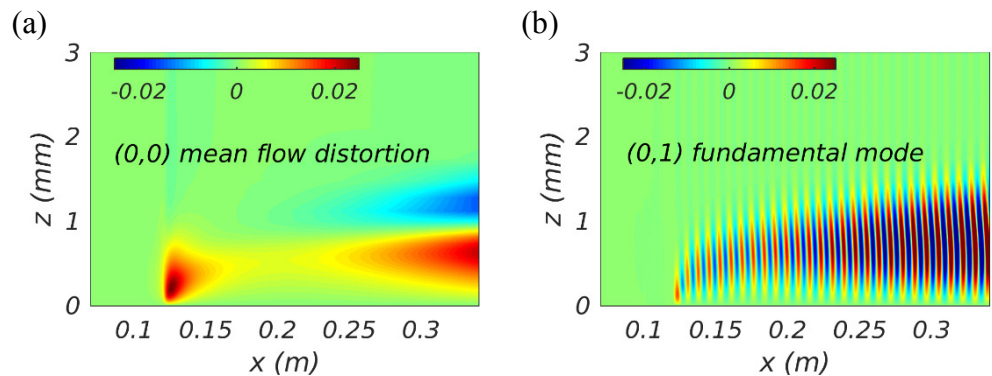

Figure 8: Stationary crossflow streamwise $u$-disturbance field generation and evolution. Plasma actuation of applied voltage of 4.0 $kV$ is concentrated at about x=0.12 $m$. (a) Shows the (0,0) mean flow distortion $u$-disturbance, while plot (b) shows the fundamental (0, 1) $u$-disturbance.

The corresponding streamwise (0,1) $u$-disturbance killer modes birth (at $x=0.12$) and subsequent continuous growth in amplitude with $x$ is shown in Figure 8 (b). The fairly rapid growth in $x$ of the (0,1) disturbance by about $x=0.25$ is then sizable enough for nonlinearity to be significant and this then induces the re-growth of the (0,0) $u$–mode (note Figure 8 (a)). That this re-growth of the mean flow disturbance is due to nonlinearity may be ascertained from Figure 9, which shows maximum amplitude evolution with $x$ of a number of higher harmonics (including the (0,0) mode) generated by the plasma field with our Navier-Stokes solver run in true linear mode for applied voltages of 3.0 $kV$ and 4.0 $kV$. Observe that the (0,0) disturbance amplitude for both voltages keeps on decreasing with $x$, downstream of the $x=0.12$ generation site. Moreover, note all the near field higher harmonic generated modes rapidly decay in magnitude too, apart from the fundamental 'killer' (0,1) mode. Finally we also show results of our linear PSE, which are in very good agreement with the LNS results once the (0,1) crossflow disturbance has established itself. Since the amplitude for the 5.0 $mm$ induced killer mode is to be used in non-linear PSE analysis to investigate the control of the most rapidly growing 7.5 $mm$ disturbance, this good agreement between PSE and LNS provides significant

confidence in the adequacy of using non-linear PSE analysis in the plasma control work presented below.

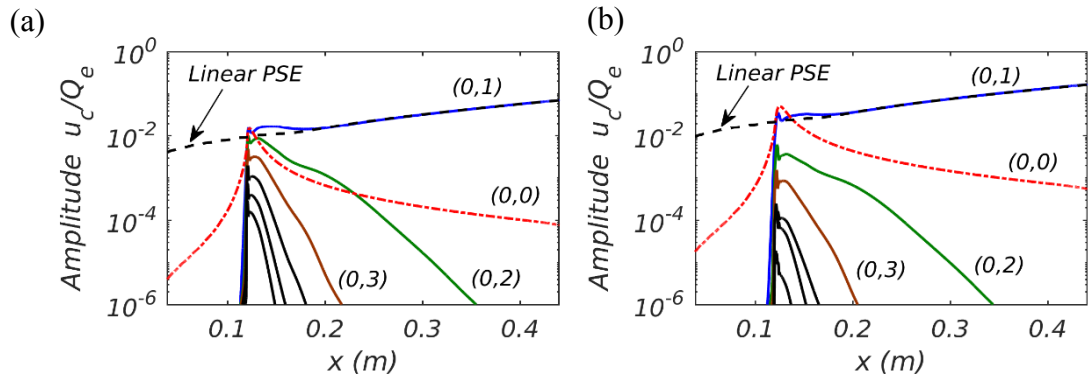

Figure 9: Left plot (a) shows 3.0 kV near field, while right plot (b) is the 4.0 kV generated near field computed with LNS, also shown is the (0, 1) linear PSE result (dashed black curves). Observe that in the 4.0 kV simulation the plasma induced mean flow correction (0,0) is much stronger and persists downstream of the plasma site.

The downstream development of the peak amplitude in the normal to leading edge component of the velocity disturbance killer mode for a range of plasma voltages is shown in Figure 10. The linear growth is seen to commence in all cases at about 200 *mm* from the leading edge, but a loss in effectiveness is also suggested with increasing applied voltage, since the change in disturbance amplitude curves going from 3.8 *kV* to 4.0 *kV* appears to be decreasing and or attaining an upper maximum limit. The development of the disturbance was compared with experimental measurements made at a plane 250 *mm* from the leading edge as shown in Figure 11, where the disturbance $u_o$ and associated dimensionalising velocity $Uo$ are defined in experimental coordinates along the freestream direction. This reveals an almost linear increase in amplitude with applied voltage for both the experiment and the numerical model with remarkably good agreement between the two. The amplitude for the highest voltage induced 5.0 *mm* killer mode was used in subsequent NPSE analysis to investigate the control of the most rapidly growing 7.5 *mm* disturbance.

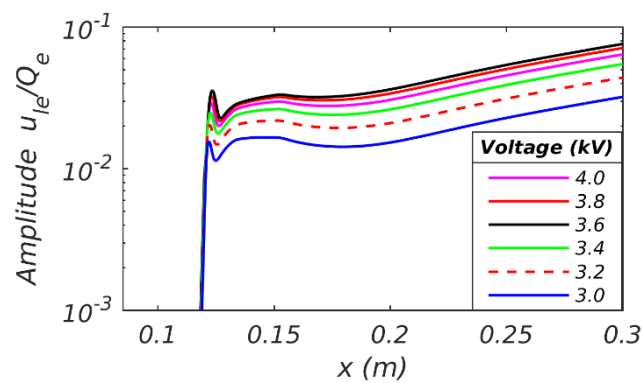

Figure 10: Profile maximum disturbance amplitude evolution of killer mode for increasing actuation voltages (3.0, 3.2, 3.4, 3.6, 3.8 and 4.0 *kV*).

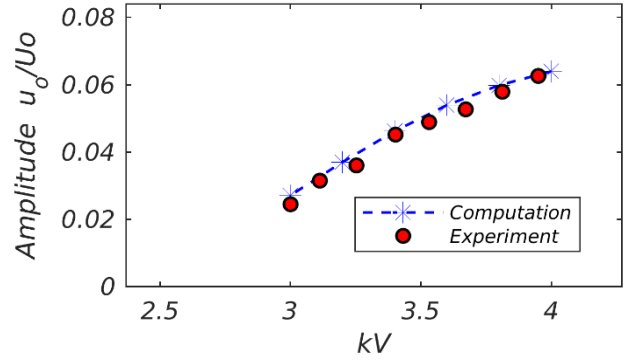

Figure 11: Amplitude of killer mode as a function of the actuator voltage on the measurement plane, 0.25 *m* from the leading edge.

The (0, 3) control mode amplitude at $x = 0.25$ *m* obtained from the LNS model with actuator voltage of 4.0 *kV* has been used to initialize an appropriate initial killer mode amplitude in a NPSE computation, the results of which are shown in Figure 12. Note from here on, we refer to the same 5.0 *mm* plasma generated, previously denoted (0,1) mode corresponding to wavenumber $\beta = 2\pi / \lambda_{5mm} = 3\beta_{PSE}$, as the (0,3) killer mode in the NPSE control analysis with a redefinition of the spanwise wavenumber $\beta_{PSE} = 2\pi / \lambda_{15mm}$. The amplitude of the transition causing target (0, 2) mode ($= 2\beta_{PSE}$) with spanwise wavelength of 7.5 *mm* was set to correspond with the experimental measurement data reported in Ustinov et al.[1] Comparison of the uncontrolled and controlled growth of the target mode indicates a clear reduction in amplitude and an expected delay in transition of about 0.1 *m*. In reality, as was mentioned previously, there was a forward movement in transition in the experiment[1] where this is explained in terms of the excitation of transition causing travelling modes by the plasma actuator due to the plasma induced unsteady velocity fluctuations which increase linearly with

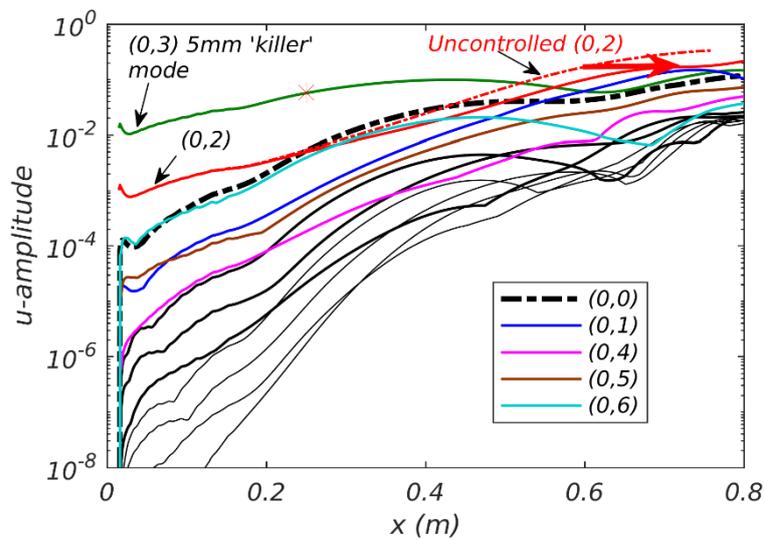

Figure 12: Non-linear PSE computation showing interaction of (0, 3) killer mode generated by plasma actuation with (0, 2) target mode for (0, 1) fundamental mode of 15 *mm*. Red star shows amplitude from LNS calculation. The *u*-amplitude is scaled with $Qe$ the local boundary-layer edge total velocity.

actuation voltage. This unsteadiness in the actuation is not captured in the actuator model described here and thus it is not possible to capture the generation of the travelling modes. However, the work does indicate that if the physical realization of the distributed plasma actuator can be redesigned so as to lessen the unsteady component of the forcing then it can be expected to be capable of producing a delay in transition due to cross-flow instability and this should be a focus for future plasma actuator development. Moreover, scaling arguments reported within the Buterfli project[28] have indicated that the approach based on distributed plasma actuation could be energetically favourable when considered for application at flight scale thus providing an important practical incentive for continued development of this technology.

# Conclusions

It has been demonstrated that the generation of "killer" modes by plasma actuation can be modelled through a linear Navier Stokes approach which when used in conjunction with a non-linear PSE solver can make a prediction about the delay in transition arising from the actuation. Although the numerical model has been shown to give good agreement with experiment for the amplitude of the killer modes on a downstream plane - the subsequent non-linear interaction leading to suppression of the transition causing "target" mode and delay in transition is not what was observed in experiment. This is due to unsteady effects in the plasma actuation and the generation of transition causing travelling crossflow modes in the experiment. Nevertheless should future refinements in the actuator hardware permit the elimination of the unsteady actuation effects, then the modelling approach outlined provides a good basis from which to predict the effectiveness of such actuators.

We have also demonstrated an efficient high fidelity LNS based methodology, which allows analysis to be undertaken relatively quickly. Comparison with linear PSE is equally good, and our LNS method may be used to provide initial amplitudes to initialize nonlinear PSE simulations. The coupled LNS-NPSE approach reproduces nearly DNS like results in fractions of the time and CPU hardware resource compared to full DNS simulations. The LNS method models the generation of crossflow disturbances, *i.e.* receptivity directly, and quite remarkable agreement with experiment has been obtained. The incorporation of body force terms within the LNS framework, we believe is ideally suited for the study of propagation, generation and control of instabilities in plasma based flow control when the instabilities to be controlled and manipulated are relatively weak.

# Appendix

**Incompressible body fitted linearised Navier Stokes disturbance equations:**
Flow variables are the three velocity terms, $(\bar{U}, \bar{V}, \bar{W})$ denoting the basic steady flow in the chordwise $(x)$, spanwise $(y)$ and wing-normal $(z)$ directions, and the associated velocity disturbance states $(u, v, w)$ and pressure disturbance $p$. Body surface curvature terms are denoted by $\kappa$ and $\chi$. Here $\kappa$ is the local body curvature while $\chi$ is given as follows

$$\chi = \frac{1}{1 - \kappa y}.$$

$$\chi u_x + w_z - \kappa\chi w + im\beta v = 0; \tag{A1a}$$

$$\frac{u_{zz} + \chi^2 u_{xx}}{R} - u\left(im\beta\bar{V} + \frac{m^2\beta^2 + \kappa^2\chi^2}{R} + \chi\bar{U}_x - \kappa\chi\bar{W}\right) - \chi P_x - \left(\bar{W} + \frac{\kappa\chi}{R}\right)u_z +$$
$$\left(\kappa\chi\bar{U} - \bar{U}_z\right)w - \chi\bar{U}u_x - \frac{2\kappa\chi^2}{R}w_x + \hat{f}_x = 0; \tag{A1b}$$

$$\frac{w_{zz} + \chi^2 w_{xx}}{R} - w\left(im\beta\bar{V} + \frac{m^2\beta^2 + \kappa^2\chi^2}{R} + \bar{W}_z\right) - P_z - \left(\bar{W} + \frac{\kappa\chi}{R}\right)w_z -$$
$$\left(2\kappa\chi\bar{U} + \chi\bar{W}_x\right)u - \chi\bar{U}w_x + \frac{2\kappa\chi^2}{R}u_x + \hat{f}_z = 0; \tag{A1c}$$

$$\frac{v_{zz} + \chi^2 v_{xx}}{R} - v\left(im\beta\bar{V} + \frac{m^2\beta^2}{R}\right) - im\beta P - \bar{V}_z w - \left(\bar{W} + \frac{\kappa\chi}{R}\right)v_z -$$
$$\chi\bar{U}v_x - \chi\bar{V}_x u + \hat{f}_y = 0. \tag{A1d}$$

In the above, $R$ is an appropriately defined Reynolds number.

## Declaration of Conflicting Interests

The Authors declare that there is no conflict of interest.

## Funding

This work was completed within the Buterfli project supported by the European Commission under the 7th Framework Programme [Grant Agreement 605605].